# Large Deflection of Cantilever Rod Pulled by Cable

Milan Batista
University of Ljubljana, Faculty of Maritime Studies and Transport
Pot pomorščakov 4, 6320 Portorož, Slovenia
milan.batista@fpp.uni-lj.si

**Abstract**

The article discusses six problems which can arise in the determination of the equilibrium configuration of an elastic cantilever rod pulled by an inextensible cable. The discussions are illustrated with graphs of equilibrium shapes and tables providing some reference numerical values.

## 1 Introduction

This paper is mainly motivated by Yau's [1] recent article in which he provided a solution of the large deflection of a guyed column pulled by an inclined cable based on an elliptic integrals description of the deformed cantilever. In particular, he considers the case when the cable length and distance of the cable support point are given and where the tension in the cable and cable inclination angle are unknowns. [We refer the reader to the Yau article for discussion of the importance of the study of cable supported structures and for some future references.] While these problems were already considered as early as the end of the 19 century by Saalschütz [2] in his book on elastic rods, where he also used elliptic integrals for the formulation of the problem, it seems that, as also observed by Yau, there are relatively few articles that consider the problem analytically.

The aim of this short article is to provide an alternative solution of the problem in terms of Jacobi elliptic functions. As Saalschütz and Yau both do we will reduce the problem to the solution of a system of two nonlinear equations; however, we will give the solutions and numerical examples of all six possible problems that can results from these equations.

## 2 Formulation of the problem

We consider an initial straight inextensible elastic cantilever rod pulled by an inextensible cable supported at points A and B where point A is the cantilever tip point and B is a fixed space point (see Figure 1). As was shown by the present author [3] the shape of the base curve of a deformed cantilever can be described by the following parametric equations

$$x(s)=\xi(s)\cos\alpha+\eta(s)\sin\alpha \quad y(s)=-\xi(s)\sin\alpha+\eta(s)\cos\alpha \quad (0\leq s\leq 1) \quad (1)$$

where

$$\xi(s)=\left[\frac{2E(k)}{K(k)}-1\right](1-s)+\frac{2}{\omega}\left[Z(\omega+K,k)-Z(\omega s+K,k)\right] \quad (2)$$

$$\eta(s)=-\frac{2k}{\omega}\left[\operatorname{cn}(\omega+K,k)-\operatorname{cn}(\omega s+K,k)\right]$$

$$\sin\frac{\alpha}{2}=k\operatorname{sn}(\omega+K,k) \quad k=\sin\frac{\varphi_0}{2} \quad (3)$$

$$\omega^2\equiv\frac{F\ell_0^2}{EI} \quad (4)$$

Here $K$ and $E$ are complete elliptic integrals of the first and second kind, $sn$, $cn$ and $Z$ are Jacobi sine, cosine and zeta elliptic functions, $F$ is applied force, $EI$ is the bending stiffness of the rod. The geometric meaning of $\xi(s)$ and $\eta(s)$ and the angles $\varphi_0$ and $\alpha$ should be clear from Figure 1. All length dimensions are normalized to cantilever length $\ell_0$.

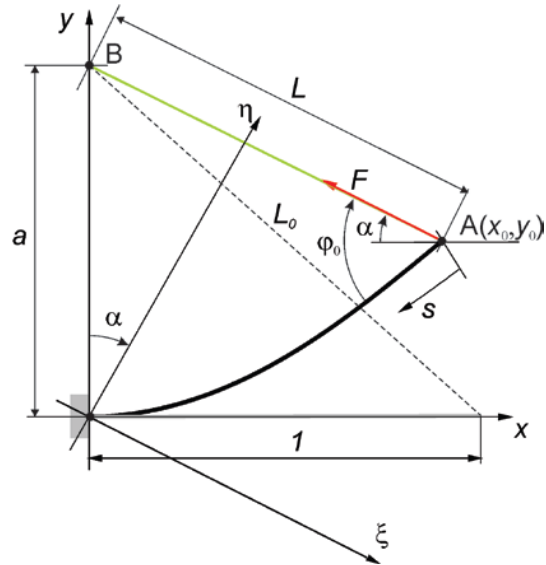

**Figure 1.** Geometry and load of the problem

As is seen from the above equations the shape of the cantilever is completely determinate when $k$ (or equivalently $\varphi_0$) and $\omega$ are known. Once they are known we can by Eq (3)$_1$ calculate $\alpha$ and further by Eqs (2) and Eqs (1) determine the cantilever shape. Two equations from which these parameters can be calculated follow from the geometry of the problem.

For given cable length $L \geq 0$ we with the help of Figure 1 obtain the following coordinates of cantilever tip point A

$$x_0 = L\cos\alpha \qquad y_0 = a - L\sin\alpha \qquad (5)$$

or in the rotated coordinate system

$$\xi_0 = L - a\sin\alpha \qquad \eta_0 = a\cos\alpha \qquad (6)$$

where $a$ is the distance between the cantilever clamped end and cable support at B.

Now, by using trigonometric identities

$$\sin\gamma = 2\sin\frac{\gamma}{2}\cos\frac{\gamma}{2} \qquad \cos\gamma = 1 - 2\sin^2\frac{\gamma}{2} \qquad (7)$$

and relation $(3)_1$ we can eliminate $\alpha$ from (6). In this way we obtain

$$\xi_0 = L - 2a\,k\,\text{sn}(\omega+K,k)\,\text{dn}(\omega+K,k) \qquad \eta_0 = a\left[1 - 2k^2\text{sn}^2(\omega+K,k)\right] \qquad (8)$$

On the other hand, the coordinates of the cantilever tip point A in the rotated coordinate system are obtained from equations (2) by setting $s = 0$

$$\xi_0 = \frac{2E(k)}{K(k)} - 1 + \frac{2}{\omega}Z(\omega+K,k) \qquad \eta_0 = -\frac{2k}{\omega}\text{cn}(\omega+K,k) \qquad (9)$$

Equating these with (8) yields

$$L - 2a\,k\,\text{sn}(\omega+K,k)\,\text{dn}(\omega+K,k) = \frac{2E}{K} - 1 + \frac{2}{\omega}Z(\omega+K,k)$$

$$a\left[1 - 2k^2\text{sn}^2(\omega+K,k)\right] = -\frac{2k}{\omega}\text{cn}(\omega+K,k) \qquad (10)$$

By assuming $1 - 2k^2\text{sn}^2(\omega+K,k) \neq 0$ and by using additional identities for elliptic functions [4] we obtain

$$a = \frac{2k\sqrt{1-k^2}}{\omega}\frac{\text{sn}(\omega,k)\,\text{dn}(\omega,k)}{2(1-k^2) - \text{dn}^2(\omega,k)} \qquad (11)$$

$$L = \frac{2E(k)}{K(k)} - 1 + \frac{2Z(\omega,k)}{\omega} + \frac{2k^2}{\omega}\frac{\text{sn}(\omega,k)\,\text{cn}(\omega,k)\,\text{dn}(\omega,k)}{2(1-k^2) - \text{dn}^2(\omega,k)} = \frac{2E(k)}{K(k)} - 1 + \frac{2Z(\omega,k)}{\omega} + \frac{ka\,\text{cn}(\omega,k)}{\sqrt{1-k^2}} \qquad (12)$$

The two relations connect four parameters: $\omega$, $k$, $a$ and $L$. We can thus choose two of them as given and the remaining two as unknowns. We thus can solve one of the six problems, which will be discussed in the next section. However, before proceeding we will present some special solutions of system (11) and (12) on the basis of the properties of Jacobi elliptic functions.

*Case when $\varphi_0 = 0$.* In this case $k = 0$. We have $E(0) = K(0) = \pi/2$ and $Z(\omega, 0) = 0$. The system is (11)-(12) therefore reduce to $a = 0$ and $L = 1$. In words: the cantilever remains straight.

*Case when $\omega = 0$.* There is no force acting on the cantilever. We have $\text{sn}(0, k) = 0$, $\text{cn}(0, k) = \text{dn}(0, k) = 1$, $\lim_{\omega \to 0} \frac{\text{sn}(\omega, k)}{\omega} = 1$ and $\lim_{\omega \to 0} \frac{Z(\omega, k)}{\omega} = 1 - \frac{E}{K}$ so the system (11)-(12) is by (3)$_2$ reduced to

$$a = \frac{2k\sqrt{1-k^2}}{(1-2k^2)} = \tan\varphi_0 \qquad L = \frac{1}{1-2k^2} = \frac{1}{\cos\varphi_0} \tag{13}$$

As expected the results correspond to the geometry of an undeformed cantilever (see Figure 1). Note that in this case $\alpha = \varphi_0$.

*Case when $\omega = 2nK$.* In this case $Z(\omega, k) = \text{sn}(\omega, k) = 0$, $\text{dn}(\omega, k) = 1$ and $\text{cn}(\omega, k) = \pm 1$ so the system (11)-(12) is reduced to

$$L = \frac{2E(k)}{K(k)} - 1 \qquad a = 0 \qquad (n = 1, 2, \ldots) \tag{14}$$

We see that both *L* and *a* are independent of the particular load parameter $\omega$. The graph of *L* as a function of $\varphi_0$ and some examples of deformed configurations of the cantilever for various *n* are shown in Figure 2. We see from the graph that $0 \leq L \leq 1$ where $L = 1$ corresponds to $\varphi_0 = 0$ and $L = 0$ to $\varphi_0 \approx \pm 0.7261661701\pi$ ($k \approx 0.9089085575$).

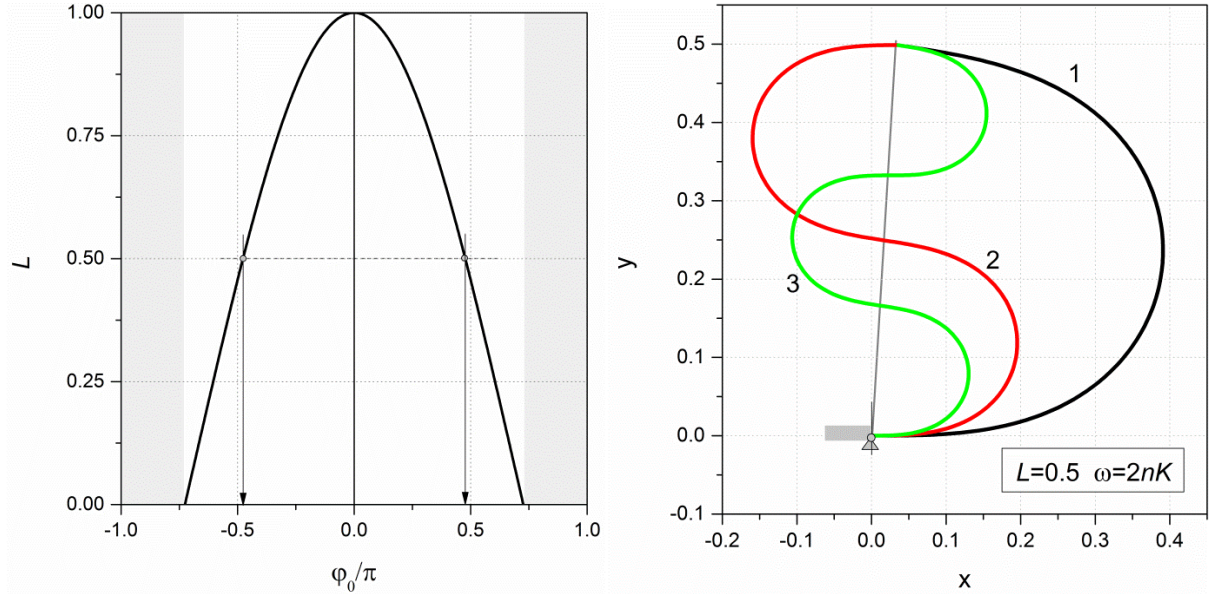

**Figure 2.** *L* as a function of $\varphi_0$ (left) and deformed cantilever shapes for the case $L=0.5$ and various values of *n*.

*Case when* $\omega=(2n-1)K$ $(n=1,2,\ldots)$. In this case $Z(\omega,k)=\operatorname{cn}(\omega,k)=0$, $\operatorname{sn}(\omega,k)=\pm 1$ and $\operatorname{dn}(\omega,k)=\sqrt{1-k^2}$. The system (11)-(12) is therefore reduced to

$$L=\frac{2E(k)}{K(k)}-1 \qquad a=\frac{(-1)^{n-1}2k}{(2n-1)K(k)} \qquad (n=1,2,\ldots) \tag{15}$$

Note that *L* and *a* are independent of load parameter $\omega$. By Eq (3)$_1$ we have $\sin\frac{\alpha}{2}=0$ so $\alpha=0$. Consequently, by Eq (5) we in this case have $x_0=L$ and $y_0=a$. In words: the cable takes a horizontal position.

In Figure 3 the graphs of *a* as a function of *L* for various values of *n* are shown. These graphs are constructed numerically from Eqs (15). Also in the same figure we present some examples of equilibrium configurations of a cantilever for the case $L=0.5$ and various values of *n*.

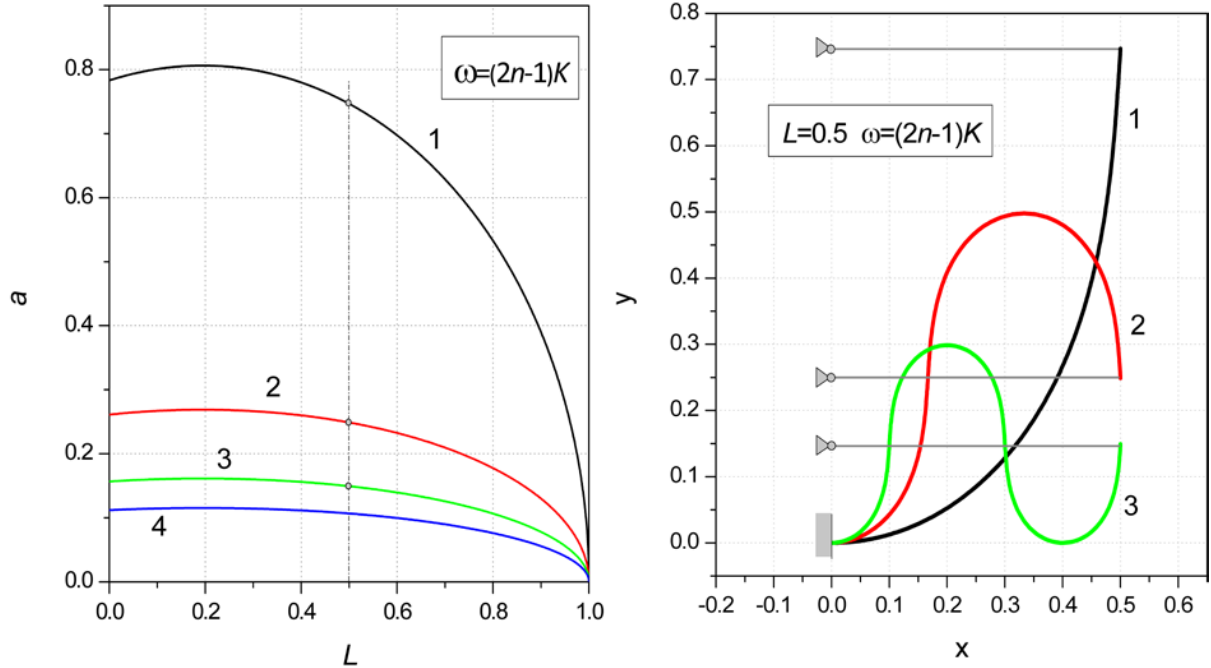

**Figure 3.** *a* as a function of *L* for $\omega = (2n-1)K$ and various values of *n* (left). Deformed cantilever for $L = 0.5$ and various values of *n* (right)

### 3 Numerical examples

In this section we will give examples of a solution of the six problems which can be derived from the basis of the system (11)-(12).

*Case when $\varphi_0$ and $\omega$ are given.* In this case *a* and *L* are explicitly given by Eqs (11) and (12). Because for the given tip angle $\varphi_0$ both *a* and *L* are one-valued functions of load parameter $\omega$ we therefore for the given data have only one possible solution. Example of graphs of *a* and *L* as a function of $\omega$ when $\varphi_0$ is given and examples of equilibrium shapes are shown in Figure 4. Some numerical values corresponding to cantilever deformed shapes in the figure are given in Table 1.

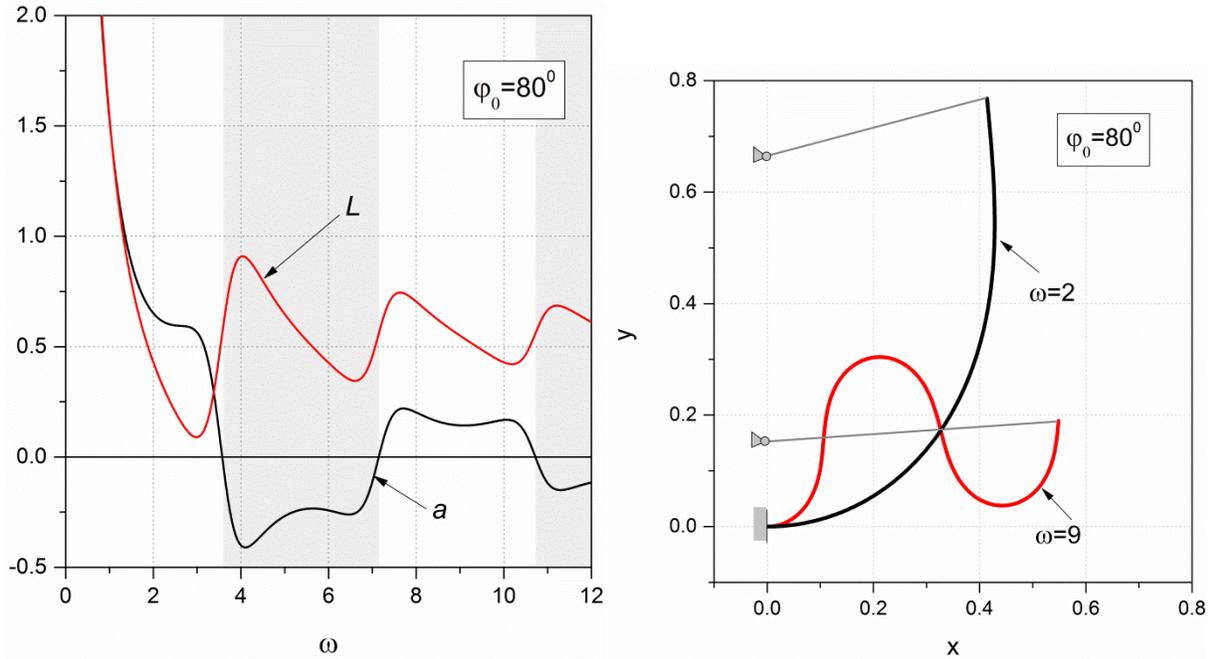

**Figure 4.** Graph of *a* and *L* as a function of load parameter $\omega$ (left) and two examples of equilibrium configurations. Shaded regions on the left correspond to negative values of *a*.

**Table 1.** Numerical values for the cantilever shapes shown in Figure 4 (right). Bold values indicate data used for calculations.

|   | $\varphi_0 \, [°]$ | $\omega$ | $a$ | $L$ | $\gamma/\pi$ | $x_0$ | $y_0$ |
|---|---|---|---|---|---|---|---|
| 1 | **80** | **2** | 0.6523121 | 0.4297663 | -0.0865992 | 0.4139593 | 0.7677970 |
| 2 | **80** | **9** | 0.1430457 | 0.5504837 | -0.0270514 | 0.5484970 | 0.1897719 |

*Case when $\omega$ and a or L are given.*  When $\omega$ and *a* are given then Eq (11) becomes a transcendental equation for an unknown *k* and Eq (12) gives *L* explicitly and when $\omega$ and *L* are given then Eq (12) becomes a transcendental equation for an unknown *k* and Eq (11) gives *a* explicitly. In both cases we have to solve a transcendental equation for an unknown *k* (or $\varphi_0$) with possible multiple solutions. The examples of these multiple solutions are shown by graphs in Figure 5 where for a given value of load parameter $\omega = 8$ we for $a = 0.2$ have three possible values of $\varphi_0$, and for the case $L = 0.8$ five possible values of $\varphi_0$. The equilibrium configurations of a cantilever corresponding to these solutions are shown in Figure 6 and some numerical values are presented in Table 2.

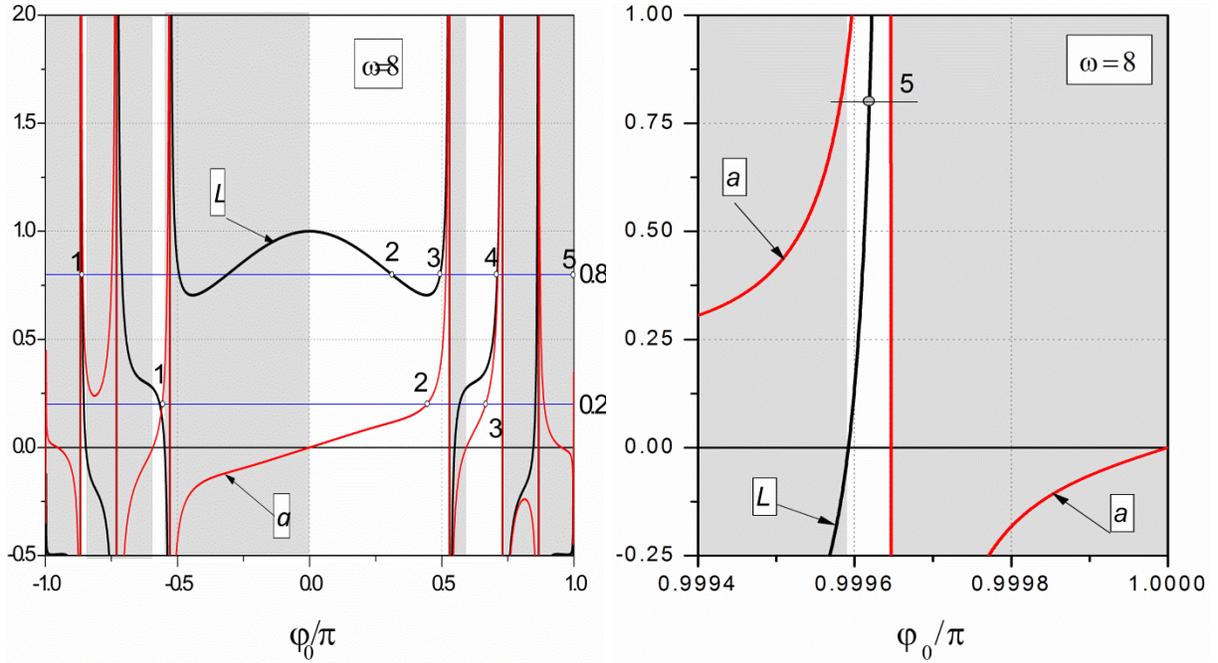

**Figure 5.** Graph of *a* and *L* for when $\omega = 8$. For the case $a = 0.2$ we have three possible solutions, for the case $L = 0.8$ we have five possible solutions.

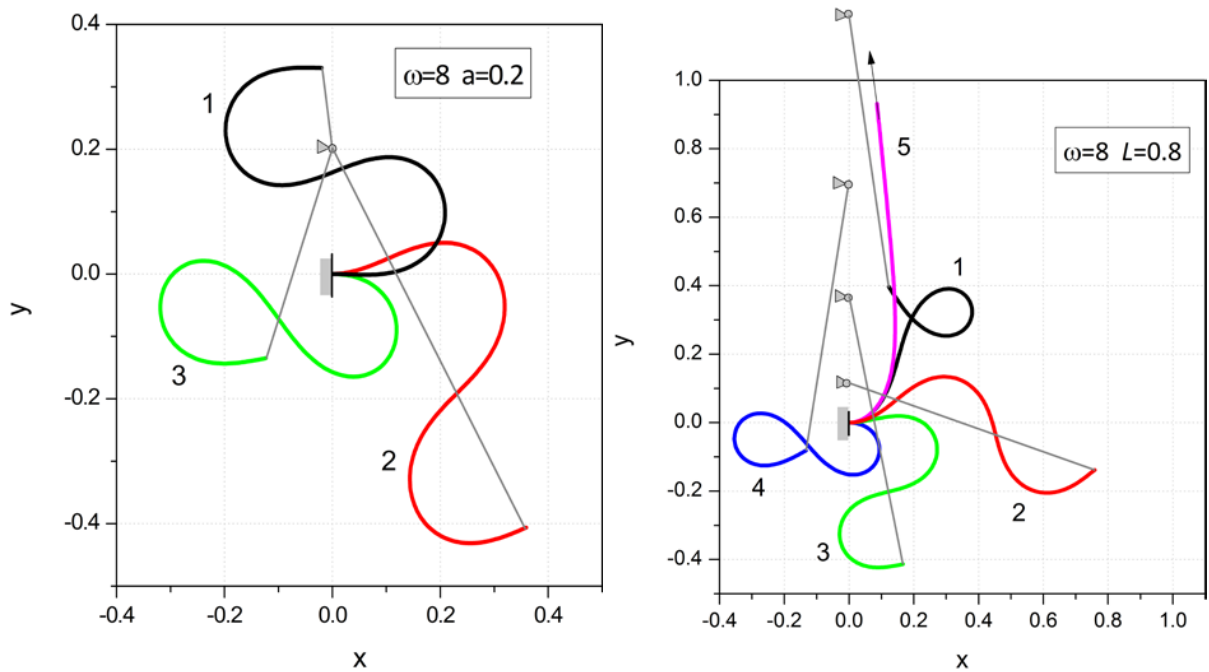

**Figure 6.** Equilibrium configurations corresponding to solutions given in Figure 5. Enumeration of shapes corresponding to cases given in Table 2.

**Table 2.** Numerical values corresponding to shapes shown in Figure 6. The bold values of *a* and *L* and $\omega = 8$ are data values.

|   | a | L | k | $\varphi_0/\pi$ | $\alpha/\pi$ | $x_0$ | $y_0$ |
|---|---|---|---|---|---|---|---|
| 1 |   | 0.1315553 | -0.7680551 | -0.5575507 | -0.5481639 | -0.0198299 | 0.3300522 |
| 2 | **0.2** | 0.7045634 | 0.6414189 | 0.4433079 | 0.3297563 | 0.3591161 | -0.4061726 |
| 3 |   | 0.3566510 | 0.8654706 | 0.6659610 | 0.6122519 | -0.1231822 | -0.1347030 |
| 1 | 1.1880104 |   | -0.9759795 | -0.8601832 | 0.4510856 | 0.1224520 | 0.3974374 |
| 2 | 0.1164301 |   | 0.4700242 | 0.3115096 | 0.1030778 | 0.7584193 | -0.1381285 |
| 3 | 0.3690474 | **0.8** | 0.7001384 | 0.4937567 | 0.4333722 | 0.1662338 | -0.4134910 |
| 4 | 0.7071851 |   | 0.8979120 | 0.7098328 | 0.5533657 | -0.1334952 | -0.0815981 |
| 5 | 1.7270341 |   | 0.9999998 | 0.9996188 | 0.4656414 | 0.0861849 | 0.9316901 |

<u>*Case when $\varphi_0$ and a or L are given.*</u> When $\varphi_0$ and *a* are given then Eq (11) becomes a transcendental equation for an unknown $\omega$ and *L* is given explicitly by Eq (12) and when $\varphi_0$ and *L* are given then Eq (12) becomes a transcendental equation for an unknown $\omega$ while *a* is explicitly given by Eq (11). In both cases we have to solve a transcendental equation for an unknown $\omega$ with - as can be concluded from the graph of *a* and *L* as functions of $\omega$ shown in Figure 4 (left) - multiple possible solutions. Since the solution of such an equation is straightforward we omit an example. However, we note that as opposed to the case when $\omega$ is given and where we can have only a finite number of possible solutions we in this case can have, when $a = 0$ or *L* is given by (14)$_1$, an infinite number of possible solutions.

<u>*Case when a and L are given.*</u> In this case the Eqs (11)-(12) become the system of two transcendental equations for unknowns $\varphi_0$ and ω. Solving the system presents no numerical difficulty; however, since the system has multiple possible solutions the contour plots of the equations like that presented in Figure 7 are helpful for providing the initial guess. An example of cantilever equilibrium configurations for two cases are presented in Figure 9. Some reference numerical values are provided in Table 3 and in Table 4 the comparison between the results for the case $a = 0.5$ and $L = 1$ given by Saalschütz [2] and the present solution are illustrated.

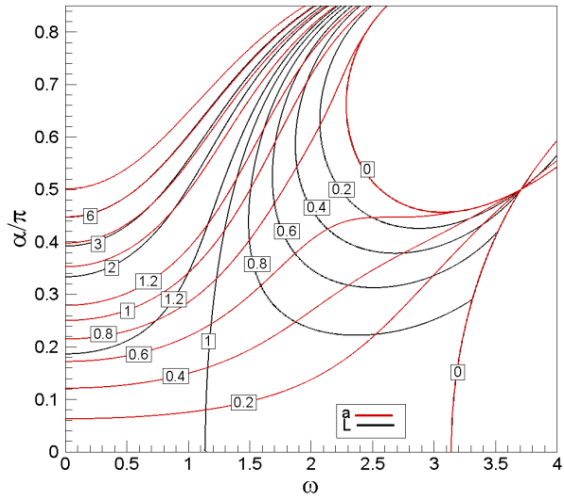

**Figure 7.** Contour plot of *a* and *L*

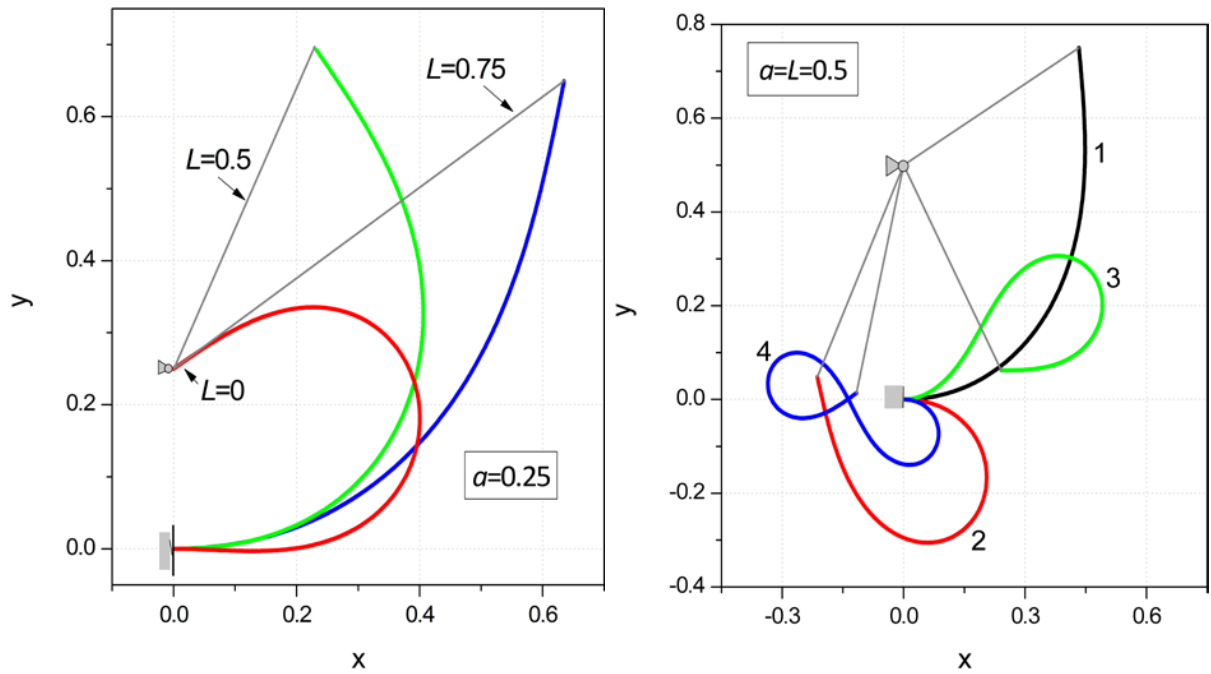

**Figure 8.** Examples of cantilever equilibrium shapes for the case $a = 0.25$ and various values of *L* (left) and the first four equilibrium configurations for the case $a = L = 0.5$ (right).

**Table 3.** Calculated values of load parameter $\omega$, angles $\varphi_0$ and $\alpha$, and free end coordinates $(x_0, y_0)$ for the case $a = L = 0.5$

| $\omega$ | $\varphi_0/\pi$ | $\alpha/\pi$ | $x_0$ | $y_0$ | Curve on Fig 8 (right) |
|---|---|---|---|---|---|
| 2.2130068 | 0.3666122 | -0.1666521 | 0.4330241 | 0.7499802 | 1 |
| 3.8999278 | -0.7811480 | 0.6403178 | -0.2133412 | 0.0477992 | 2 |
| 5.7452174 | -0.6585928 | 0.3397495 | 0.2412215 | 0.0620363 | 3 |
| 8.4495375 | 0.7554032 | 0.5749502 | -0.1166467 | 0.0137968 | 4 |

**Table 4.** Comparison of results for the case $L = 1$ and $a = 0.5$

| | $x_0$ | $y_0$ | $\dfrac{4\omega^2}{\pi^2}$ | $\alpha$ | $\varphi_0 - \alpha$ |
|---|---|---|---|---|---|
| Saalschütz [2] pp 165 | 0.96 | 0.23 | 0.5420 | $15^0 40'$ | $20^0 32'$ |
| Present | 0.9649 | 0.2376 | 0.556818 | $15^0 12' 54''$ | $21^0 10' 16''$ |

For practical purposes it is interesting to know the relation between load parameter $\omega$ and cable length for a given *a*. To obtain such dependence the system (11)-(12) was for a given *a* repeatedly solved with decreasing length *L* starting with initial length $L_0 = \sqrt{1 + a^2}$ and the initial guess $\tan\varphi_0 = a$ and $\omega = 0$. In order to compare load parameter $\omega$ for different values of *a* we instead of length *L* use cable relative length defined as

$$\varepsilon = \frac{L_0 - L}{L_0} \tag{16}$$

The graph of $\omega$ as a function of $\varepsilon$ for various values of *a* is shown in Figure 9. In the figure we can observe a strong nonlinear dependence between $\omega$ and $\varepsilon$. We note also that when $a \geq 1$ the relative length has the upper limit

$$\lim_{\omega \to \infty} \varepsilon = 1 - \frac{a - 1}{\sqrt{a^2 + 1}} \tag{17}$$

Practically, this limit can never be reached since the cantilever can never be completely directed vertically.

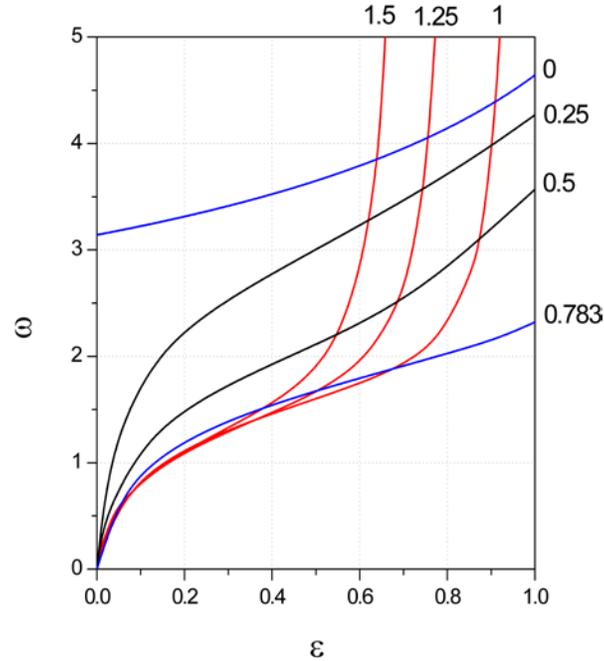

**Figure 9.** Dependence of load parameter $\omega$ on relative cable length $\varepsilon = 1 - L/L_0$ for various values of *a*.

## 4 Conclusions

We presented a new solution of the problem of determining the equilibrium configurations of a cantilever pulled by cable. The problems were formulated as two relations connecting load parameter $\omega$ with geometry parameters *a, L* and $\varphi_0$. We show that the obtained relations lead to six possible problems, where only the problem with a given $\omega$ and $\varphi_0$ has a unique solution. All other pairs of data yield multiple possible solutions. The numerical values given for various types of problems can be used as reference values for solutions obtained by other methods.

**References**

1. Yau, J.D., *Closed-Form Solutions of Large Deflection for a Guyed Cantilever Column Pulled by an Inclination Cable.* Journal of Marine Science and Technology-Taiwan, 2010. **18**(1): p. 130-136.
2. Saalschütz, L., *Der belastete Stab unter Einwirkung einer seitlichen Kraft*. 1880: B. G. Teubner, Leipzig.
3. Batista, M., *Equilibrium Configurations of Cantilever under Terminal Loads.* arXiv:1303.6490v2 [physics.gen-ph], 2013.
4. Olver, F.W.J. and National Institute of Standards and Technology (U.S.), *NIST handbook of mathematical functions*. 2010, Cambridge: Cambridge University Press. xv, 951 p.